\documentclass[journal]{IEEEtran}

\usepackage{ifpdf}
\usepackage{cite}
\usepackage{amsmath,amssymb,amsfonts}
\usepackage{algorithmic}
\usepackage{graphicx, adjustbox}
\usepackage{array}
\usepackage{enumitem}
\usepackage{breqn}
\usepackage{ dsfont }
\usepackage{ upgreek }
\usepackage{textcomp}
\usepackage{multirow}
\usepackage{algorithm}
\usepackage{algorithmic}
\usepackage{caption}
\usepackage{pgfplots}
\usepackage{tkz-euclide}
\usepackage{subcaption}
\usepackage{tikz}{\tiny }
\usepackage{hyperref}
\usepackage{balance}
\begin{document}

\title{Privacy-Preserved Task Offloading in Mobile Blockchain with Deep Reinforcement Learning}

	\author{Dinh C. Nguyen,~\IEEEmembership{Student Member,~IEEE,}
	Pubudu N. Pathirana,~\IEEEmembership{Senior Member,~IEEE,} \\
	Ming Ding,~\IEEEmembership{Senior Member,~IEEE,}
	Aruna Seneviratne,~\IEEEmembership{Senior Member,~IEEE}
	
	\thanks {*This work was supported in part by the CSIRO Data61, Australia.}
	\thanks{Dinh C. Nguyen is with School of Engineering, Deakin University, Waurn Ponds, VIC 3216, Australia, and also with the Data61, CSIRO, Docklands, Melbourne, Australia  (e-mail: cdnguyen@deakin.edu.au).}
	\thanks{Pubudu N. Pathirana is with School of Engineering, Deakin University, Waurn Ponds, VIC 3216, Australia (email: pubudu.pathirana@deakin.edu.au).}
	\thanks{Ming Ding is with Data61, CSIRO, Australia (email: ming.ding@data61.csiro.au).}
	\thanks{Aruna Seneviratne is with School of Electrical Engineering and Telecommunications, University of New South Wales (UNSW), NSW, Australia (email: a.seneviratne@unsw.edu.au).}
}

\markboth{IEEE Transactions on Network and Service Management}%
{Shell \MakeLowercase{\textit{et al.}}: Bare Demo of IEEEtran.cls for IEEE Journals}

\maketitle

\begin{abstract}
Blockchain technology with its secure, transparent and decentralized nature has been recently employed in many mobile applications. However, the process of executing extensive tasks such as computation-intensive data applications and blockchain mining requires high computational and storage capability of mobile devices, which would hinder blockchain applications in mobile systems. To meet this challenge, we propose a mobile edge computing (MEC) based blockchain network where multi-mobile users (MUs) act as miners to {offload their data processing tasks and mining tasks} to a nearby MEC server via wireless channels. Specially, we formulate task offloading, user privacy preservation and mining profit as a joint optimization problem which is modelled as a Markov decision process, where our objective is to minimize the long-term system offloading utility and maximize the privacy levels for all blockchain users. We first propose a reinforcement learning (RL)-based offloading scheme which enables MUs to make optimal offloading decisions based on blockchain transaction states, wireless channel qualities between MUs and MEC server and {user's power hash states}. To further improve the offloading performances for larger-scale blockchain scenarios, we then develop a deep RL algorithm by using deep Q-network which can efficiently solve large state space without any prior knowledge of the system dynamics. Experiment and simulation results show that the proposed RL-based offloading schemes significantly enhance user privacy, and reduce the energy consumption as well as computation latency with minimum offloading costs in comparison with the benchmark offloading schemes.
\end{abstract}


\IEEEpeerreviewmaketitle

\section{Introduction}
\label{sec:introduction}
In recent years, the blockchain technology has been employed widely in various industrial applications such as Internet of Things (IoT), healthcare, industrial applications, etc. \cite{1}, \cite{2}. Blockchain works as a peer-to-peer public ledger where users can store data (i.e., records of transactions) and share information with other blockchain nodes in a trustworthy and decentralized manner. With the advancement of mobile technologies, blockchain now can be implemented in mobile devices to provide more flexible blockchain-based solutions for IoT applications \cite{3}, \cite{4}. The foundation of the efficient and secure operation of blockchain is a computation process known as \textit{mining} \cite{5}. In order to append a new transaction to the blockchain, a blockchain user, or a miner, needs to run a mining puzzle, i.e., Proof of Work (PoW) or Proof of Stake (PoS) which is generally complicated and requires intensive computations. Resource-limited IoT nodes or mobile devices, therefore, cannot participate in the mining operation, which can restrict the application of blockchain in mobile systems. Recently, mobile edge computing (MEC), which enables mobile devices to offload their computation tasks to a nearby computationally powerful MEC server, provides effective solutions to solve computation issues of mobile devices \cite{6,7,8}. By employing highly computational MEC resources, the performance of MEC-based mobile systems may be improved significantly, such as reducing computation latency and improving computation efficiency \cite{9,10,11}.  

{In the literature, to solve the offloading issues in mobile systems, various approaches have been proposed, using convex optimization-based methods \cite{12}, \cite{13}, or machine learning techniques such as reinforcement learning (RL) \cite{14,15,16} or deep reinforcement learning (DRL) \cite{17,18,19,20}. In mobile blockchain applications, MEC-based computation offloading strategies to optimize mining services \cite{21} have been considered in some previous studies \cite{22,23,24}.} Such pioneering studies consider offloading issues, from data offloading, resource allocation and block size optimization to enhance offloading efficiency in blockchain networks. Addition to computation offloading, privacy is also another critical issue that should be considered carefully when designing offloading systems. Offloading data tasks and mining tasks to MEC server to reduce computing burden on mobile devices while ensuring high privacy for mobile users is very important in network management and ensures the robustness of the blockchain-based systems. However, we find that the study that can achieve optimal policies of offloading mining tasks with enhancing privacy awareness in mobile blockchain is still missing. Many pioneering blockchain schemes \cite{22,23,24}  have concentrated on only offloading of mining tasks without considering user privacy preservation, which has been one of the key challenges in MEC-based blockchain networks. 

To this end, we propose a DRL-based dynamic task offloading scheme for a MEC blockchain network where each mobile user (MU) acts as a miner that {can offload its IoT data processing tasks and mining tasks} to the MEC server for computing services. Particularly, we formulate task offloading and user privacy preservation as a joint optimization problem. Then we develop a RL-based algorithm to solve the proposed optimization problem with a simplified blockchain model. Next, to break the curse of high dimensionality in state space when increasing the number of blockchain users, a DRL-based method called deep Q-network (DQN) algorithm is proposed. The objective of the proposed scheme is to obtain optimal offloading actions for all blockchain miners such that the user privacy level is maximized while minimizing computation latency and energy consumption for offloading. It is worth noting that in this paper, we only concern about the optimization of the offloading decision of each user, while edge resource is assumed to be sufficient to perform all offloading tasks. The main differences between our study and the existing schemes \cite{21,22,23,24} are highlighted as follows.

\begin{enumerate}
	\item We consider a new task offloading model for both IoT data processing tasks and mining tasks in a MEC-enabled blockchain network.  Mobile users can act as miners to learn dynamic computation offloading policies to perform IoT data processing and participate in the mining puzzle to make extra profits.	
	\item  	To obtain optimal offloading policies for all miners, we propose a RL-based algorithm and then use a deep Q network to implement our proposed offloading strategies, aiming to achieve the best privacy performance and mining profits while minimizing offloading latency and energy cost. 
   \item We conduct {both experiments and numerical simulations} to verify the advantages of the proposed offloading algorithm over the other offloading schemes.
\end{enumerate}

The remainder of this paper is organized as follows. Section~\ref{RelatedWorks} reviews the related MEC-blockchain works. Section~\ref{Section:SystemModel} introduces the system model, and then we present the formulation of computation model in Section~\ref{Section:SystemFormulation}. Based on system formulation, in Section~\ref{Section:Offloading_Optimization} we derive the offloading optimization problem with a RL-based solution. Next, we propose two offloading schemes in Section~\ref{Section:Algorithms} by using RL and DRL. Experiment and simulation results are given in Section~\ref{Section:Simulation}. Finally, our conclusions are drawn in Section~\ref{Section:Conclusions}.
\section{Related Works}
\label{RelatedWorks}
The works in \cite{12}, \cite{13} were introduced with the objective of minimizing the energy usage or task execution latency for edge task offloading by leveraging traditional convex optimization methods, but they usually require prior system knowledge. RL has emerged as a strong alternative that can derive an optimal solution via trial and error without requiring any prior environment information \cite{14}. Nevertheless, in high-dimensional offloading problems, the dimension of state and action space can be extremely high that makes RL-based solutions inefficient \cite{15},  \cite{16}. Fortunately, DRL methods \cite{17} such as DQN have been introduced as a strong alternative to deal with such high-dimensional problems and demonstrates its scalability and offloading efficiency in various MEC-based applications, such as multi-base station virtual MEC \cite{18}, multi-user MEC system \cite{19}, and multi-IoT networks \cite{20}, \cite{21}.

Further, a computation offloading scheme was proposed in \cite{22} for a wireless blockchain networks where mobile users can offload their blockchain mining tasks to a MEC server or to a network of device-to-device users, subject to probabilistic backhaul and latency constraints.  Meanwhile, the problem of computation power allocation for mining task offloading in the MEC-enabled blockchain was investigated in \cite{23} with the main focus on reward optimization for mobile terminals. Further, the work \cite{24} considered an optimization problem of offloading scheduling, resource allocation and adaptive block size scheme for a blockchain-based video streaming system with MEC. The user targets to offload computation to MEC nodes or mobile users such that transcoding revenue is maximized while achieving the high latency efficiency.

In addition to mining efficiency, privacy of blockchain users is another important issue that needs to be considered for mobile offloading in MEC-based blockchain networks \cite{25,26,27}. However, most existing computation offloading frameworks for blockchain mining services \cite{22,23,24} have ignored user privacy.  A user privacy model was proposed for MEC-based mobile networks where mobile devices can select efficient offloading decisions using a constrained Markov decision process (CMDP) \cite{29}. Meanwhile, the work in \cite{28} proposed a privacy-aware computation offloading scheme using reinforcement learning, which enables IoT devices to learn an offloading policy so as to protect their personal information while achieving minimum system utility loss. 

\section{System Model}
\label{Section:SystemModel}
\subsection	{Blockchain Network}
\begin{figure}
	\centering
	\includegraphics[width=0.95\linewidth]{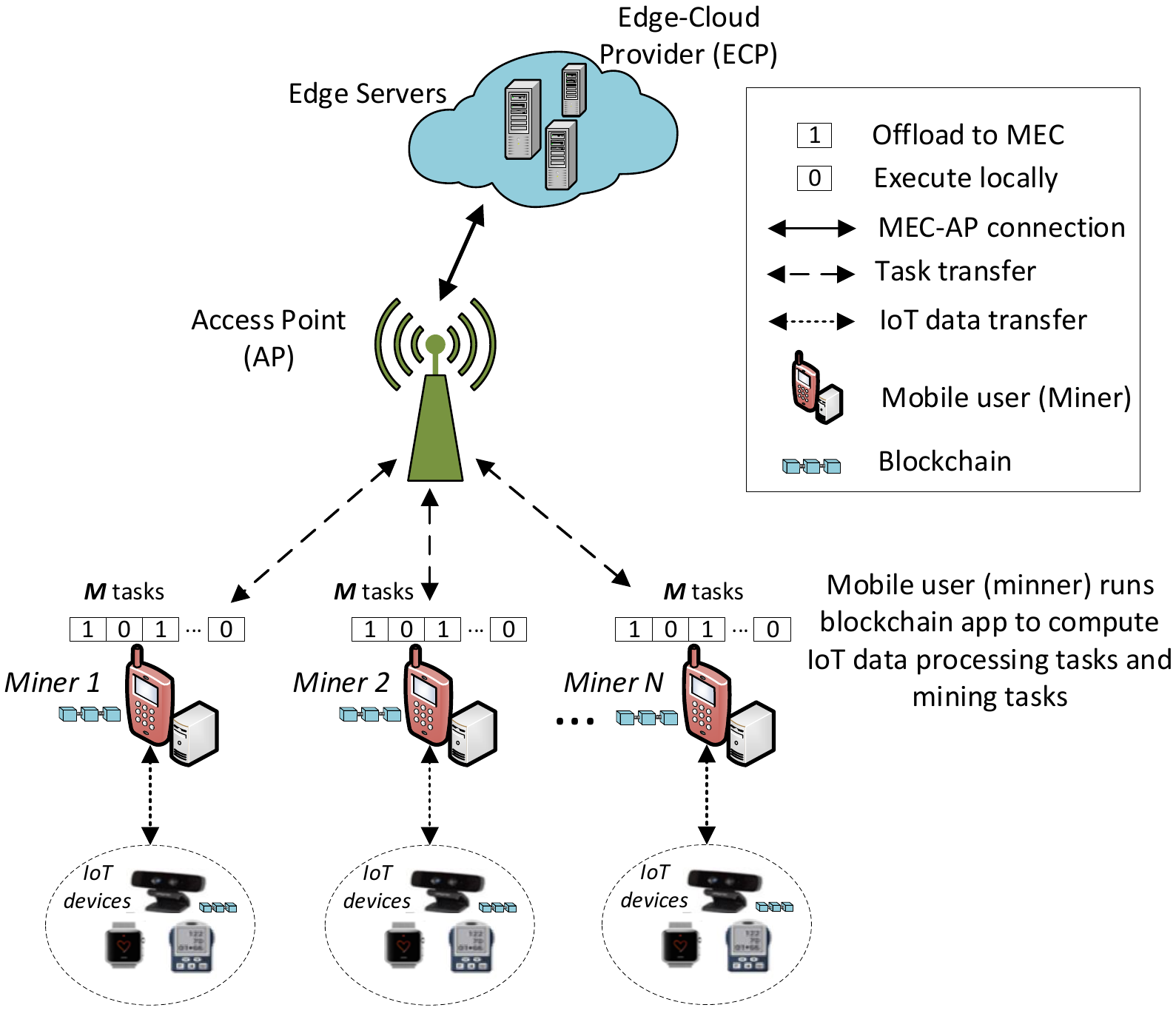}
	\caption{The proposed mobile edge blockchain architecture. }
	\label{Overview}
	\vspace{-0.15in}
\end{figure}

In this paper, we consider a public blockchain network for MEC-enabled IoTs as shown in Fig.~\ref{Overview}. {The blockchain network consists of the MEC server, a wireless access point (AP), a set of multiple IoT devices and \textit{N} MUs denoted by a set $\mathcal{N} = \{1,2,..., N\}$. Blockchain is used as decentralized database to record IoT transactions and manage securely data exchange between IoT devices, MUs and MEC server. In our blockchain scenario, each IoT device is regarded as a lightweight node which is responsible to perform data collection from external sensing environments (e.g., wearable sensors for sensing health data \cite{2}) and transmit wirelessly such data to MUs for necessary data processing. It is assumed that IoT devices are grouped and managed by a certain MU in a local private network. Specially, MUs represent their IoT devices to perform transactions with the MEC-blockchain network and execute sensing data. Here, each MU performs two computation functionalities for: IoT data processing tasks and mining tasks. For IoT data processing tasks, MUs can decide to execute locally or offload tasks to the MEC server for efficient computation. Meanwhile, for mining tasks, MUs can act as miners and purchase hash power from the edge cloud providers to perform mining and make extra profits \cite{30}.} (MU, miner and mobile miner terms are used interchangeably throughout the paper). In this way, miners can gather transactions of the blockchain network and group them into data blocks, which are then verified and appended into the blockchain after consensus verification. The miner which is the first one to solve the consensus puzzle is rewarded for their mining contribution, and the verified transaction is broadcast to other blockchain nodes in secure and immutable ways.

Without loss of generality, we consider that each MU \textit{n} has multiple IoT data processing tasks denoted by a set $\mathcal{M} = \{1,2,..., M\}$. Therefore, we denote $R_{nm}$ as the \textit{m-th} data task of user n. Each miner is identified by a unique blockchain account. We also assume that the MEC server has sufficient computation resources with high-frequency CPU cores and storage capacity, and can provide computation services for a certain number of mobile miners in the blockchain network.

\begin{figure}
		\centering
	\includegraphics[width=0.95\linewidth]{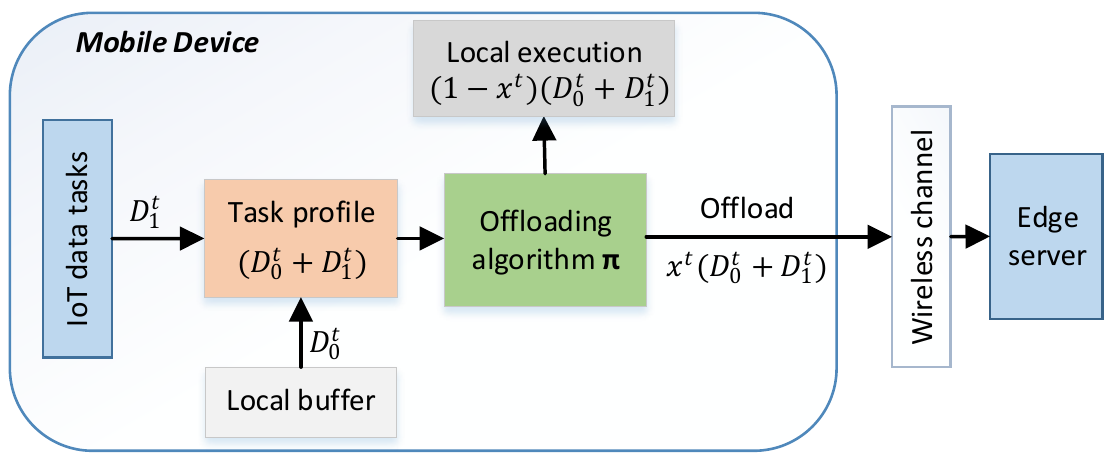}
	\caption{The offloading procedure in mobile edge blockchain. }
	\label{offloading_procedure}
	\vspace{-0.15in}
\end{figure}

It is also assumed that time is slotted and at each time slot \textit{t}, the miner \textit{n} generates a task $R_{nm}$. For each miner \textit{n}, the data processing task at the timeslot \textit{t} can be formulated as a variable tuple $R_{nm}^t \triangleq (D_{1nm}^t, D_{0nm}^t, X_{nm}^t, \tau_{nm}^t) \in \mathcal{R}$. Here $D_{1nm}^t$ (in bits) denotes the data size of IoT data newly received from IoT devices by the miner \textit{n} at the timeslot \textit{t}. $D_{0nm}^t$ (in bits) is the data size of IoT data in the buffer of the miner \textit{n}.  $X_{nm}^t$ (in CPU cycles/bit) denotes the total number of CPU cycles required to accomplish the computation for the task $R_{nm}^t$. Moreover, $\tau_{nm}^t$ (in seconds) reflects the maximum tolerable delay of task $R_{nm}^t $.

\subsection{{Offloading For IoT Data Processing Tasks}}
In this paper, we consider a realistic blockchain scenario where the transactions from IoT devices are highly dynamic. That means at a certain time, the transaction volume can be very large (i.e., during the data collection of IoT devices) and at another time, this can be very small (i.e., idle mode of IoT devices). Based on that, at the timeslot $t$, if the size of data processing task is too large, the miners have to divide the current IoT data into two parts: data processing task for execution and the rest for storage in the buffer for future process. The mining strategy with task offloading is shown in Fig.~\ref{offloading_procedure} and can be explained as follows. 

At each time slot $t$, the miner receives new IoT data $D_{1nm}^t$ from IoT devices and adds it into a block based on the blockchain concept \cite{1}. Besides, the miner also has to process the previous IoT data $D_{0nm}^t$ stored in the buffer. Moreover, for simplicity, we assume that each block contains only a transaction which will be verified by the miner through a mining process (e.g., PoW) so that the verified transaction can be appended to blockchain. As a result, the total blockchain data of a transaction at each time slot $t$ is ($D_{0nm}^t+D_{1nm}^t$) which can be formulated as a data processing task $R_{nm}^t $. The miner will choose to offload the data processing task to the MEC server or execute it locally and store the rest data processing tasks into the local buffer for processing in the future. Accordingly, we consider two computing modes in this paper, namely:
\begin{itemize}
	\item \textit{Offloading computing}: The miner \textit{n} offloads its data processing tasks to the nearby MEC server.
	\item \textit{Local computing}: The miner \textit{n} executes its data processing tasks at the local mobile devices. 
\end{itemize}
\begin{table}
	\caption{List of notations. }
	\scriptsize
	\centering
	\captionsetup{font=scriptsize}
	\setlength{\tabcolsep}{5pt}
	\begin{tabular}{|c|p{6cm}|}
		\hline
		\textbf{Notation}& 
		\textbf{Description}
		\\
		\hline
		$D_{1nm}^t$& Blockchain transaction for task $m$ at miner $n$ at time slot $t$
		\\
		\hline
		$D_{0nm}^t$& Blockchain transaction in the buffer of miner
		\\
		\hline
		$ X_{nm}^t$ & The total number of CPU cycles per task
		\\
		\hline
		$\tau_{nm}^t $& Completion deadline for mining a task
		\\
		\hline
		$x_{nm}^t$& The task offloading policy
		\\
		\hline
		$g^t$& Wireless channel power gain
		\\
		\hline
		$f_{nm}$& MEC computational capacity
		\\
		\hline
		$L_{nm}^{o,t}$& Latency of edge computation 
		\\
		\hline
		$E_{nm}^{o,t}$& Energy consumption of edge computation
		\\
		\hline
		$L_{nm}^{l,t}$& Latency of local computation 
		\\
		\hline
		$E_{nm}^{l,t}$& Energy consumption of local computation 
		\\
		\hline
		$P_{nm}^{t}$& The total task offloading privacy level
		\\
		\hline
		$p_n$& The miner $n$'s hash power purchased from edge-cloud provider
		\\
		\hline
		$R_n^{mining}$& The mining reward of miner $n$
		\\
		\hline
		{$\alpha$}&	{ The learning rate of deep Q-learning}
			\\
		\hline
			{$\gamma$ }& 	{The discount factor of deep Q-learning}
			\\
		\hline
		{ $\mathcal{D}$ }& 	{The replay memory capacity}
			\\
		\hline
	\end{tabular}
	\label{Table:notation}
	\vspace{-0.15in}
\end{table}

Specially, we introduce an offloading decision policy denoted by a binary variable $x_{nm}^t \in \{0,1\} $. Here $x_{nm}^t = 1$ means that the miner \textit{n} offloads the task \textit{m} to the MEC server over the wireless channel with the wireless channel power gain $g(t)$, and $x_{nm}^t = 0$ means that the miner \textit{n} decides to process the task \textit{m} locally. For simplicity, the wireless channel power gain $\{g^t\}_{t\geq0}$ can be formulated as a Markov chain model with $Pr(g^{t+1} =m|g^t=n) = h_{nm}, \forall n,m \in \mathcal{G}$, where $\mathcal{G}$ is the wireless channel state set. Particularly, we define that $g^t=1$ or $(g^t=0)$ represents the good (or bad) channel state at the timeslot $t$. This policy will be used to analyze channel conditions for ensuring user privacy during task offloading. The notations used in this paper are summarized in Table~\ref{Table:notation}.

\subsection{{Offloading For Blockchain Mining Tasks}}
In addition to IoT data computation, MUs can participate in the consensus process to obtain rewards as their mining contribution. {However, due to the high computation requirement by the mining puzzle and resource constraints of mobile devices, cloud-based mining methodologies such as the leased hashing power technique \cite{30} can be adopted which enables MUs to lease the required amount of hashing power from edge cloud servers}. In this way, MU can purchase the necessary hash power from the edge cloud providers (e.g., AWS cloud services) for mining blocks. From the user point of view, the cloud-based mining process can eliminate any hassle of managing the infrastructure, i.e, installing extra mining hardware or requiring more computing resources on mobile devices. Furthermore, MUs can gain more economic benefits for their mobile blockchain apps through mining rewards, which will attract participation of more users and in return improve the robustness of the IoT blockchain network. {Note that in our concerned blockchain network, MUs can leverage existing mining communication protocols such as Stratum \cite{31} to achieve low-latency and secure mining process}. 

\subsection{Offloading Privacy}
In the task offloading process, the MEC server is assumed to be curious about the personal information of miners. Motivated by \cite{28}, in this paper, we consider two privacy issues in the task offloading process, namely \textit{user location privacy} and \textit{usage pattern privacy} of the miner. By monitoring task offloading history of the miner, the MEC server can infer channel state records to obtain location information. Further, the MEC server can infer personal information of the miner by tracking mining data through the task offloading period, which may raise a concern of privacy threat based on the usage pattern of the miner. Thus, how to build an optimal task offloading policy that improves user privacy while guaranteeing low computation costs in the dynamic wireless environment is important to any MEC-based mobile blockchain applications. In the next subsection, we present the offloading problem formulation for the proposed model.

\section{Problem Formulation}
\label{Section:SystemFormulation}
In this section, we formulate the offloading problems for IoT data tasks and data processing tasks in details.

\subsection{{Offloading Cost of IoT Data Processing Tasks}}
In this subsection, we derive the expression of performance metrics on network latency, energy consumption and user privacy under two offloading cost categories. 
\subsubsection{Task offloading cost}
We model the case when the MU \textit{n} selects to offload its data processing tasks to the MEC server ($x_{nm} =1$). Offloading latency and energy consumption problems are considered in our offloading formulation.

- \textit{Offloading latency:} The computation latency is the key performance indicator for evaluating the quality of data processing task offloading. For the miner \textit{n}, ($ \forall \textit{n} \in \mathcal{N}$), the whole computation latency to offload the data processing task \textit{m} to the MEC server at the timeslot \textit{t} can be decomposed into three parts, namely uploading, queuing and  processing \cite{11}, which is given as
\begin{equation}
L_{nm}^{o,t} = L_{nm}^{u,t} + L_{nm}^{p,t} + L_{nm}^{q,t},
\end{equation}
where $L_{nm}^{u,t}, L_{nm}^{p,t}, L_{nm}^{q,t}$ are {uploading delay, task processing delay and queuing latency} on MEC server, respectively.

- \textit{Uploading delay:} 
We consider the delay when the user \textit{n} uploads the data processing task \textit{m} to the MEC server. We use $r_n$ to denote the upload transmission rate of the user \textit{n} in the wireless channel. For simplicity, we assume that the transmission rate is the same for all users \textit{n}. Thus, we can express the required time to upload the data processing task \textit{m} to the MEC server as
\begin{equation}
L_{nm}^{u,t} = \dfrac{D_{0nm}^t+D_{1nm}^t}{r_n}.
\end{equation}

- \textit{Queuing latency:} 
In our task offloading model, we are interested in the queuing latency caused by waiting in the task buffer for processing at the MEC server. Let $Q$ denote as the total number of CPU cycles in the data processing task buffer, the queuing latency can be given by
\begin{equation}
L_{nm}^{q,t} = \dfrac{Q}{f_{nm}}. 
\end{equation}

- \textit{Processing delay:} 
The time consumed by the miner \textit{n} to execute the data processing task \textit{m} on the MEC server at the timeslot \textit{t} can be formulated as \cite{22}
\begin{equation}
L_{nm}^{p,t} = \dfrac{(D_{0nm}^t+D_{1nm}^t)X_{nm}^t}{f_{nm}}, 
\end{equation}
where $f_{nm}$ is defined as the computational resource (in CPU cycles per second) allocated by the MEC server to accomplish the task $R_{nm}$. Note that this computation capacity is assumed to be large enough to serve all miners our blockchain system. 

Once the execution of data processing task on MEC server finishes, the processed result is downloaded to the miner. Thus the downloading time latency can be given by $L_{nm}^d = \dfrac{G_{nm}}{S_{nm}}  $ where $G_{nm}$ defines the size of executed data and $S_{nm}$ is the data rate of the MU \textit{n}. However, the processed data from the MEC server is very small and the download rate is very high in general \cite{12}, and thus the latency (and energy cost) for the download process will be ignored in this paper.

- \textit{Energy consumption:} The entire energy consumption for the task offloading can be formulated as the sum of energy costs for task uploading, task processing and MEC operation. Thus it is easy to express the total energy consumed by the computation offloading for the data processing task \textit{m} of the MU \textit{n} at the time slot \textit{t} as
\begin{equation}
E_{nm}^{o,t} = P^ML_{nm}^{u,t} + \gamma(f_{nm})^3L_{nm}^{p,t} + P^CL_{nm}^{q,t},  
\end{equation}
where $\gamma$ is the energy consumption efficiency coefficient of the MEC server, $P^M$ is the transmit power of miner and $P^C$ denotes the baseline circuit power \cite{13}. 

\subsubsection{Local computation cost}
When the MU \textit{n} decides to execute its  task \textit{m} locally ($x_{nm} =0$), it uses computation resource of the local device to process the mining puzzle. In this case, we consider the time cost and energy consumption for the local mining process.

We denote the local executing time per data bit as $t_{nm}^l$ (in sec/bit), then the total time cost consumed by the miner \textit{n} to complete the task \textit{m} at the timeslot \textit{t} can be expressed by
\begin{equation}
L_{nm}^{l,t} = (D_{0nm}^t+D_{1nm}^t)t_{nm}^l.
\end{equation}
Meanwhile, we use $e_{nm}^l$ (in J/bit) to denote the energy consumption per data bit of the MU \textit{n}. Accordingly, the energy cost required for the MU \textit{n} to execute the data processing task at the timeslot \textit{t} is
\begin{equation}
E_{nm}^{l,t} = (D_{0nm}^t+D_{1nm}^t)e_{nm}^l.
\end{equation}
\subsection{User Privacy}

In this subsection, we consider user privacy issues that have been largely ignored in previous studies in MEC-based mobile blockchain networks. Here, offloading privacy is defined as the preservation of the user information (including usage pattern and user location) during the mobile task offloading process against unintended usage or threats. Through the user's wireless transmission activity, an eavesdropper can monitor the service migrations among MEC and users and track physical movements of the user. An attacker can estimate the size of the sensing data newly generated and thus evaluate the usage pattern if the user offloads all the sensing data to the edge device under good radio channel state. {Furthermore, the user also should prevent himself from being tracked by the eavesdropper or malicious MEC servers, and the offloading strategy needs to mislead the attackers to avoid any privacy harms on the user \cite{37}.} Thus, user information can be securely maintained which improves significantly user privacy, accordingly. Based on above discussion, here we consider two privacy issues for task offloading: usage pattern privacy and location privacy. 

\subsubsection{Usage pattern privacy} Naturally, if the wireless channel state is good, the miner tends to offload their data processing task with all blockchain transaction data ($ D_{offload}^t = D_{1nm}^t+D_{0nm}^t$) to the MEC server to reduce processing time and energy cost on the mobile device. More specially, if the mobile user \textit{n} moves near to the access point, it is more likely to have $ D_{offload}^t = D_{1nm}^t$ (due to $D_{0nm}^t=0$). Obviously, the data usage pattern of the miner can be obtained by the MEC server through monitoring the data processing task $D_{offload}^t$. We denote $\zeta$ as the pre-defined good channel power gain state, similar to \cite{28}, the level of usage pattern privacy can be estimated by
\begin{equation}
P_{nm}^{u,t} =|D_{0nm}^t-x_{nm}^t(D_{0nm}^t+D_{1nm}^t)|.\mathds{I}(g_n^t \ge {\zeta}),
\end{equation}
where $\mathds{I}$ denotes the indicated function that equals 1 (or 0) if the statement is true (or false). The equation (8) means that the miner \textit{n} deliberately changes the amount of transaction data processed by local device under the good channel state, aiming to create a difference between the actual transaction data size $D_{0nm}^t$ and the offloading data size ($D_{0nm}^t+D_{1nm}^t$) to preserve usage pattern privacy.

\subsubsection{Location privacy} Besides usage pattern privacy, the location privacy is another critical issue that should be considered to improve the performance of blockchain data processing task offloading. According to the system model formulation in the previous subsection, the MU will offload its data processing task to the MEC server if the wireless channel has a good transmission state ($g_{nm}^t=1$). Otherwise, the data processing task will be executed locally or stored in the buffer for future process when the channel state is bad ($g_{nm}^t=0$). In fact, we can encourage more mobile users to offload tasks to the MEC server when the channel state is good for efficient computation. However, since the wireless channel power gain is highly correlated to the distance between the miner and MEC server, this offloading strategy can expose location information of miner \cite{37}. Applying the privacy metric in \cite{28}, we can formulate the location privacy level as
\begin{equation}
P_{nm}^{l,t} =\mathds{I}[ x_{nm}^t(D_{0nm}^t+D_{1nm}^t)].\mathds{I}(g_n^t < \zeta),
\end{equation}
which means that the miner should keep offloading its data processing task to MEC server under poor transmission channel quality to preserve its location privacy. In summary, to protect privacy during task offloading, miners should mitigate the offloading rate under good channel quality while increasing the offloading rate under poor channel quality. 

Combining the result (8) and (9), the total task offloading privacy level at the timeslot \textit{t} can be given as
\begin{equation}
\label{Equation:Privacy}
P_{nm}^{t} = P_{nm}^{u,t}+\lambda P_{nm}^{l,t},
\end{equation}
where $\lambda $ scales the importance of the location privacy relative to the usage pattern privacy ($ 0<\lambda<1$).

\subsection{{Reward of Blockchain Mining}}
{As part of the blockchain network, MUs can join the consensus process to perform mining tasks for extra profits. In the blockchain network, MUs compete against each other to become the first one to solve the mining puzzle. Each MUs $n$ determines its mining service demand to purchase the hash power from the edge-cloud provider, denoted as $p_n$ (Hash/sec). Then, the relative hash power of the MU $n$ to the blockchain network can be defined as }
\begin{equation}
\upmu_n = \dfrac{p_n}{H},
\end{equation}
where $H$ represents the total hash power of the blockchain network which can be estimated through the compact status reports from miners \cite{32}. It can be seen that when $\upmu_n$ increases, the probability to achieve successful consensus increases, which will increase the mining reward for the MU $n$, accordingly.

In general, in order to mine successfully a block, two steps are needed including the mining step and the propagation step. In the mining step, the probability that the MU $n$ mines the block is proportional to its relative hash power $\upmu_n$. The mined block is then propagated to the blockchain network. However, there is possibility that this MU $n$ propagates the mined block slower with other miners in the propagation step, which makes such a block likely to be discarded from the blockchain. This issue is called as orphaning \cite{33}, and this miner does not receive a reward for its mining. According to \cite{34}, the orphaning probability is approximated as $\mathcal{P}_{orphan} = 1- e^{-\upeta\upphi(s_n)}$, where $\upeta$ is a constant mean value and set $\upeta =1/600(sec)$ \cite{35}. Further, $s_n$ denotes the number of transactions contained in the block that is mined by the miner $n$, and $\upphi(s_n)$ is a function of block size \cite{36}, which represents the block propagation time. Clearly, with a larger block size $s_n$, the propagation time required to reach consensus of a block will be larger. As a result, the probability of successful mining by miner $n$ can be expressed as 
\begin{equation}
\mathcal{P}= \upmu_n(1- \mathcal{P}_{orphan}) = \upmu_ne^{-\upeta\upphi(s_n)}. 
\end{equation}

We denote $\mathcal{R}$ as the reward of the first miner which achieves consensus, then the expected mining reward of the MU $n$ can be calculated as $R_n = \mathcal{R}\mathcal{P} = \mathcal{R}\upmu_ne^{-\upeta\upphi(s_n)}$. Moreover, the process of solving the mining puzzle incurs an associated cost, i.e., a payment from the miner $n$ to the edge-cloud provider, denoted as $Y_n$ (token). Finally, based on the above mining formulation, the total mining reward of data processing tasks of the miner $n$ can be expressed as
{\begin{equation}
	\label{Equation:MiningReward}
	R_n^{mining} = \mathcal{R}\upmu_ne^{-\upeta\upphi(s_n)} - Y_n.
	\end{equation}}

\section{Offloading Optimization with RL}
\label{Section:Offloading_Optimization}
In this section, we derive the optimization problem and then formulate a RL-based approach for the proposed scheme.
\subsection{Optimization Problem Formulation}
In this section, we formulate the data processing task offloading and edge resource allocation as a joint optimization problem. The objective in this work is to maximize the offloading privacy while minimizing the sum cost of computation latency and energy consumption. 

We first formulate the computation latency as the maximum of the local task processing time and the task execution time on MEC server, that can be expressed as
	\begin{equation}
	L_{nm}^{t} = \max \{L_{nm}^{o,t},L_{nm}^{l,t}\}.
	\end{equation}

Meanwhile, the total energy consumption includes the local energy consumption $E_{nm}^{l,t}$ and energy cost for the taks offloading $E_{nm}^{o,t}$. Therefore, we can denote $C$ as the cost function that is the weighted sum of the time latency and energy consumption, as
\begin{equation}
\label{Equation:SystemCost}
C_{nm}^{t} = \sum_{n=1}^{N}\sum_{m=1}^{M}[\alpha_1(x_{nm}^tE_{nm}^{o,t}+(1-x_{nm}^t)E_{nm}^{l,t}) + \alpha_2.L_{nm}^{t}],
\end{equation}
where $ \alpha_1, \alpha_2 \in (0, 1)$ denote the  weight of energy consumption and task processing latency. Now we can formulate the optimization problem to jointly optimize the system privacy~\ref{Equation:Privacy}, system cost~\ref{Equation:SystemCost}, and mining reward~\ref{Equation:MiningReward} under the constraint of maximum task mining latency and MEC computation capacity, which can be expressed as follows
\begin{align}
&P1: \notag 
&\max_{\substack{\textbf{x}}}
\sum_{n=1}^{N}\sum_{m=1}^{M} (P_{nm}+R_n^{mining}-C_{nm} ) 
\end{align}
\begin{subequations}
subject to
	\begin{align}
	&x_{nm} \in \{0,1\}, \forall \textit{n} \in \mathcal{N},  \textit{m} \in \mathcal{M}, \label{C1}
	\\
	& x_{nm}L_{nm}^o+(1-x_{nm}L_{nm}^l) \leq \tau_{nm}, \forall \textit{n} \in \mathcal{N},  \textit{m} \in \mathcal{M}.  \label{C2}
	\end{align}
\end{subequations}

Here, $\textbf{x} = [x_{11}, x_{12},...,x_{1M};...;x_{N1}, x_{N2},...,x_{NM}]$ is the offloading decision vector. Here, the constraint~\ref{C1} represents the binary offloading decision policy of the miner \textit{n} for the task m, or offloading to the MEC server or processing locally at the mobile device. Further, the execution time to complete a data processing task should not exceed a maximum time latency value, which is expressed in the constraint~\ref{C2}. It is worth noting that the optimization problem $(P1)$ is a mixed integer non-convex problem, which is NP-hard and difficult to derive an optimal solution. This is because the variable $x_{nm} $ is binary variables which make the problem $(P1)$ become a mixed integer programming problem that is in general non-convex and NP-hard. Further, the feasible set of problem $(P1)$ is also not convex. Also, the objective function is non-convex due to the product relationship between offloading decisions, the energy consumption factor and the offloading latency factor. To solve the offloading issues in the proposed task offloading, there are some challenges that need to be solved: 
	\begin{itemize}
	\item 	The system states and rewards cannot be obtained in advance by step-wise control. Thus, the conventional optimization methods that only consider the current state may not be feasible. 
	\item 	The joint optimization problem of offloading decision, offloading cost and mining profit with dynamicity of offloading data sizes, wireless channel states and hash power states is of high dimensionality and high complexity. It is hard to make joint efficient decisions by conventional approaches. 
	\item 	As shown in the previous work \cite{1}, there are some regularities of user's preferences and network features. With these regularities, offloading strategies and mining strategies can be made in advance. Therefore, using deep reinforcement learning approach would be a viable solution to address the above challenges.
\end{itemize} 

To overcome such challenges, we propose a dynamic offloading scheme using deep reinforcement learning (DRL). The advantages of our algorithm are twofold. First, it enables miners to obtain an optimal offloading action at each system state based on current blockchain transaction data and channel state without requiring prior knowledge of system dynamics. Second, as the DRL-based method is efficient in solving complex problems with large state space \cite{18}, it can achieve a better offloading performance to improve the quality of large-scale blockchain applications. Details of our design are presented in the next section. 

\subsection{Reinforcement Learning Formulation}
\begin{figure}
	\centering
	\includegraphics[width=0.95\linewidth]{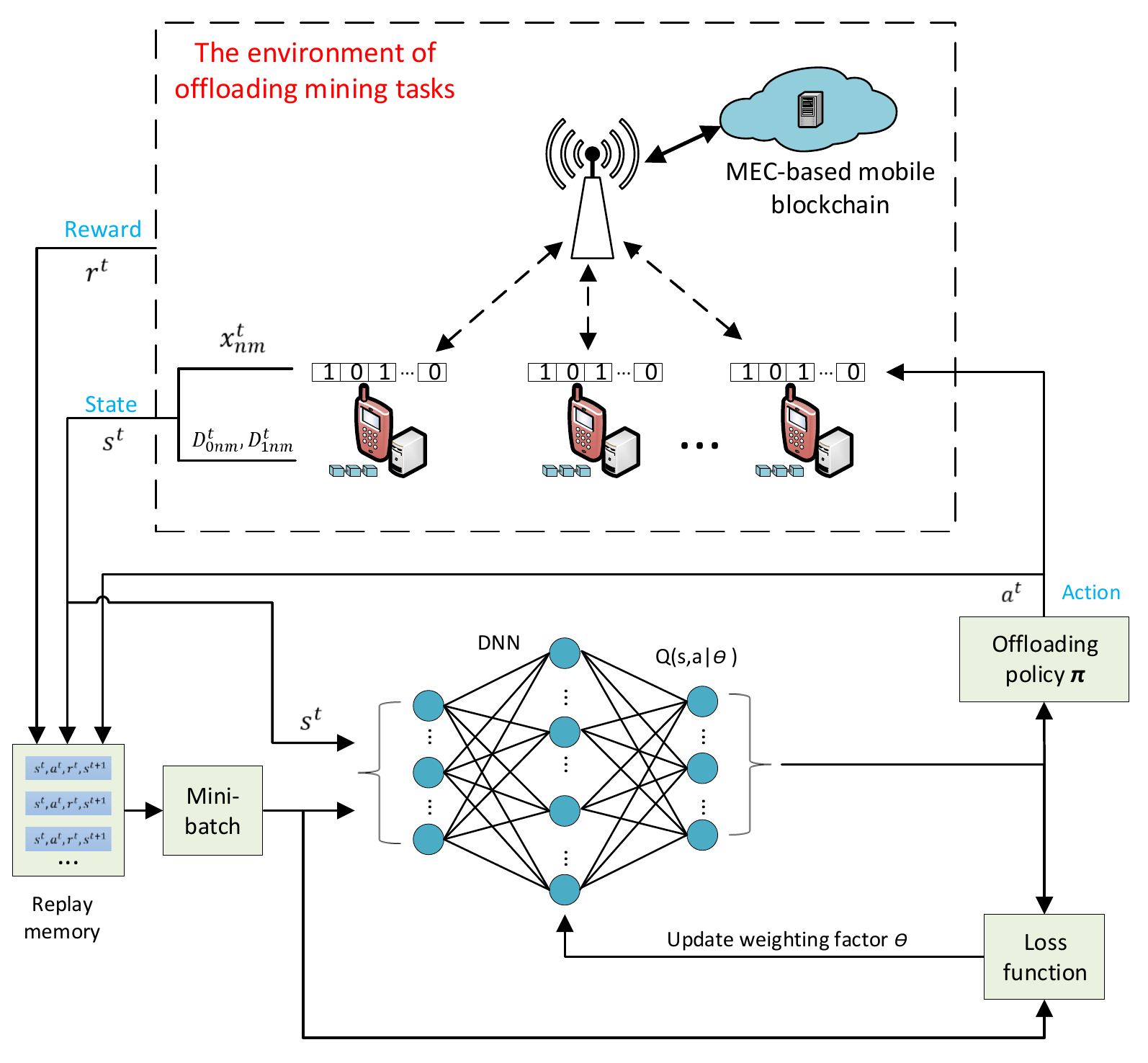}
	\caption{ RL-based offloading for mobile blockchain. }
	\label{RLOffloading}
	\vspace{-0.15in}
\end{figure}
In our blockchain scenario, each miner acts an agent which interacts closely with the mobile blockchain environment at the timeslot \textit{t} to find a task offloading action $a$ for a state $s$ using a policy $\pi$ as shown in Fig.~\ref{RLOffloading}. This policy is defined as a mapping from the action to the state, i.e., $\pi(s)=a$. The main goal of the blockchain miner is to find an optimal policy $\pi$, aiming to maximize the total amount of award $r$ over the long run. To implement the RL-based algorithms, we first define the specific state, action and reward for the proposed task offloading.
We formulate the main elements of a reinforcement learning approach for our blockchain offloading as follows. 

\subsubsection{State} The system state is chosen as $s^t= \{D_{1nm}^t, D_{0nm}^t, g_n^t, p_n^t, Y_n^t\}$ where $D_{1nm}^t, D_{0nm}^t$ represent the new and buffered blockchain transaction data of the miner at the timeslot \textit{t}, respectively. $g_n^t$ is the  power gain state of wireless channel between the miner and MEC server as defined in the system model. Further, $p_n^t$ represents the hash power that the miner $n$ purchases from the edge-cloud provider, and  $Y_n^t$ is the payment that the miner $n$ has to pay for the edge-cloud provider to perform mining.
\subsubsection{Action} In our paper, power gain was pre-defined using the channel state transition probability, and the MEC computation capacity is assumed to be large enough to serve all miners in our blockchain system. For hash power and payment, each user determines its mining service demand to purchase the hash power from the edge-cloud provider before the data offloading starts, while payment is completed after the data offloading process in an automatic manner. As a result, in this paper we only need to consider the offloading policy of data tasks as the action space. It can be formulated as the offloading decision vector $\textbf{x}^t = [x_{11}^t, x_{12}^t,...,x_{1M}^t;...; x_{N1}^t, x_{N2}^t,...,x_{NM}^t]$. Therefore, the action vector can be expressed as $a^t = {x}_{nm}^t (	\forall \textit{n} \in \mathcal{N},  \textit{m} \in \mathcal{M})$.
\subsubsection{Reward} The objective of the RL agent is to find an optimal offloading decision action $a$ at each state $s$ with the aim of achieving the maximum mining reward $R(s,a)^{mining}$ and the highest privacy level $P(s,a)$ while minimizing the sum cost $C(s,a)$ of time and energy consumption in the task offloading. Specially, the reward function should be positively related to the objective function of the optimization problem (P1) in the previous section. Accordingly, we can formulate the immediate system reward $r^t(s,a)$ as
\begin{equation}
\label{Equation:Reward}
r^t(s,a) = P^t(s,a) +R_n^{mining}- C^t(s,a). 
\end{equation}
In the next section, we propose task offloading schemes using Reinforcement Learning with two algorithms: RL-based task offloading (RLO) and deep RL-based task offloading (DRLO).
\section{RL-based Algorithms}
\label{Section:Algorithms}
\subsection{RL-based Task Offloading Algorithm} 
The principle of Reinforcement Learning (RL) can be described as a Markov Decision Process (MDP) \cite{15}. In the RL model, an agent can make optimal actions by interacting with the environment without an explicit model of the system dynamics. In our blockchain scenario, we consider miners as agents to develop the RL scheme. At the beginning, the miner has no experience and information about the blockchain environment. Thus it needs to \textit{explore} for every time epoch by taking some actions at each offloading state, i.e., the size of current blockchain transaction data. As long as the miner has some experiences from actual interactions with the environment, it will \textit{exploit} the known information of states while keep exploration. In this paper, as a combination of Monte Carlo method and dynamic programing, a temporal-difference (TD) approach is employed to allow the agent to learn offloading policies without requiring the state transition probability which is difficult to acquire in realistic scenarios like in our dynamic mobile blockchain. In this subsection, we focus on developing a dynamic offloading scheme using a free-model RL. To this end, the state-action function can be updated using the experience tuple of agent $(s^t,a^t, r^t, s^{t+1})$ at each time step t as
\begin{equation}
\label{Equation:Qlearning}
Q(s^t,a^t) \leftarrow Q(s^t,a^t) + \alpha \sigma^t,
\end{equation}
which is called as Q-learning algorithm \cite{15}. Here,  $\sigma^t = r(s^t,a^t) + \gamma*\max Q(s^{t+1},a^{t+1}) - Q(s^t,a^t)$ is the TD error which will be zero for the optimal Q-value. Further, $\alpha$ is the learning rate, $\gamma$ is the discount factor between (0,1). In line with the discussion, under the optimal policy $\pi^*$ which can be obtained from the maximum Q-value ($\pi^*(s) = arg maxQ^*(s,a)$), the Bellman optimality equation \cite{15} for the state-action equation can be expressed as 
\begin{equation}
Q^*(s^t,a^t) = \mathds{E}_{s^{t+1} \sim E}[r(s^t,a^t) + \gamma*\max Q^*(s^{t+1},a^{t+1})].
\end{equation}
It is noting that the Q-learning algorithm is proved to converge with probability one over an infinite number of times \cite{15} and achieves the optimal $Q^*$. Specially, we focus on finding the optimal policy $\pi^*$ which maximizes the reward $r(s,a)$ in equation~\ref{Equation:Reward}. The details of task offloading implementation for mobile blockchain using Q-learning (RLO) are summarized in Algorithm~\ref{Algorithm_RL}. 

\begin{algorithm}
	\caption{RL-based task offloading (RLO) algorithm for mobile blockchain networks}
	\label{Algorithm_RL}
	\begin{algorithmic}[1]
		\STATE \textbf{Initialization:}
		\STATE Initialize parameters: learning rate $\alpha$, discount factor $\gamma$, exploration rate $\epsilon \in (0,1)$
		\STATE Initialize the action-value function Q with initial pair $(s,a)$, and create a Q-table with $Q(s^0,a^0)$
		\STATE Set $t=1$
		\STATE \textbf{Procedure:}
		\WHILE {$t \leq T$} 
		\STATE \textit{/$***$ Plan the task offloading in blockchain $***$/}
		\STATE Observe blockchain transaction ($D_1^t,D_0^t$)
		\STATE Estimate the channel gain state $g^t$
		\STATE Set $s^t = \{D_1^t,D_0^t, g^t \}$
		\STATE Select a random action $a^t$ with probability $\epsilon$, otherwise $a^t = argmaxQ(s^t,a,\theta)$
		\STATE Offload data processing task $x^t(D_1^t+D_0^t)$ to MEC server or execute  $(1-x^t)(D_1^t+D_0^1)$ locally
		\STATE Calculate the system reward $r^t$ by equation~\ref{Equation:Reward}
		\STATE Estimate the privacy level $P(s,a)^t$
		\STATE Estimate the system cost $C(s,a)^t$
		\STATE \textit{/$***$ Update $***$/}
		\STATE Set $s^{t+1} = \{D_1^{t+1},D_0^{t+1}, g^{t+1} \}$
		\STATE Update $Q(s^t,a^t)$ by equation~\ref{Equation:Qlearning}
		\STATE $t \leftarrow t+1$
		\ENDWHILE
	\end{algorithmic}
	
\end{algorithm}

Using Algorithm~\ref{Algorithm_RL}, the optimization problem (P1) can be solved to obtain the optimal offloading policy for miners. The RLO algorithm consists of two phases, the planning phase and the updating phase. The inputs are system states, i.e blockchain transaction data and channel states, and actions, i.e offloading tasks to MEC server or executing locally. The outputs are the resulting $Q^t(s,a)$ with maximum reward $r^*(s,a)$ which corresponds to the offloading policy $\pi^*(s,t)$ in each state. In the planning phase (lines 6-15), we use a $\epsilon$-greedy policy to balance the exploration and exploitation \cite{15} for updating the Q function in equation~\ref{Equation:Qlearning}. At each time epoch, the miner observes the blockchain state, selects an offloading action, and estimates the privacy value and system costs. After each action, the miner moves to the next step, updates the new state (lines 17-19) and iterates the offloading algorithm to obtain the optimal offloading policy. 

Although the RLO algorithm can solve the problem (P1) by obtaining the optimum reward, there are still some remaining problems. The state and action values in the Q-learning method are stored in a two-dimensional Q table, but this method can become infeasible to solve complex problems with a much larger state-action space. This is because if we keep all Q-values in a table, the matrix $Q(s,a)$ can be very large, which makes the learning agents difficult to obtain sufficient samples to explore each state, leading to the failure of the learning algorithm. Moreover, the algorithm will converge very slow due to too many states that the agent has to process. To overcome such challenges, in the next subsection, we propose to use deep learning with Deep Neural Network (DNN) to approximate the Q-values instead of using the conventional Q-table.
\subsection{DRL-based Task Offloading Algorithm} 
In the DRL-based algorithm, a DNN is used to approximate the Q-values $Q(s^t,a,\theta) $ with weights $\theta$ as shown in Fig.~\ref{RLOffloading}. Further, to solve the instability of Q-network due to function approximation, the experience replay solution is employed in the training phase with the buffer $\mathcal{B}$ which stores experiences $e^t= (s^t,a^t,r^t,s^{t+1})$ at each time step $t$. Next, a random mini-batch of transitions from the replay memory is selected to train the Q-network. Here the Q-network is trained by iteratively updating the weights $\theta$ to minimize the loss function, which is written as
\begin{equation}
L^t(\theta^t) = \mathds{E}  [(r^t +\gamma*\max Q(s^{t+1},a'|\theta')-Q(s^t,a^t|\theta^t))^2],
\end{equation}
where the $\mathds{E}[]$  denotes the expectation function. The detailed DQN-based task offloading algorithm for the proposed blockchain network is summarized in Algorithm~\ref{Algorithm_DRL}. 
\begin{algorithm}
	\caption{DRL-based task offloading (DRLO) algorithm for mobile blockchain networks}
	\label{Algorithm_DRL}
	\begin{algorithmic}[1]
		\STATE \textbf{Initialization:}
		\STATE Set replay memory $\mathcal{D}$ with capacity $N$
		\STATE Initialize the Q network with input pair $(s,a)$ and estimated action-value function $Q$ with random weight $\theta$;
		initialize the exploration probability $\epsilon \in (0,1)$
		\FOR{$t = 1,2,...$}
		\STATE \textit{/$***$ Plan the task offloading in blockchain $***$/}
		\STATE Observe blockchain transaction ($D_1^t,D_0^t$)
		\STATE Estimate the channel gain state $g^t$
		\STATE Set $s^t = \{D_1^t,D_0^t, g^t \}$
		\STATE Select a random action $a^t$ with probability $\epsilon$, otherwise $a^t = argmaxQ(s^t,a,\theta)$
		\STATE Offload data processing task $x^t(D_1^t+D_0^t)$ to MEC server or execute  $(1-x^t)(D_1^t+D_0^1)$ locally
		\STATE Observe the reward $r^t$ calculated via equation~\ref{Equation:Reward} and  next state $s^{t+1}$
		\STATE Evaluate the achieved privacy $P(s,a)^t$, and system cost $C(s,a)^t$
		\STATE \textit{/$***$ Update $***$/}
		\STATE Store the experience ($s^t,a^t,r^t,s^{t+1}$) into the memory $\mathcal{D}$
		\STATE Sample random mini-batch of state transitions $s^t,a^t,r^t,s^{t+1}$ from  $\mathcal{D}$
		\STATE Calculate the Q-value by $y^k= r^k + \gamma*\max Q(s^{k+1},a', \theta)$
		\STATE Perform a gradient descent step on $(y^k-Q(s^k,a^k,\theta))^2$
		as loss function
		\STATE Train the Q-network with updated $\theta^t$
		\ENDFOR 
	\end{algorithmic}
	
\end{algorithm}

The DRLO algorithm can achieve the efficient task offloading strategy in an iterative manner. As shown in Algorithm~\ref{Algorithm_DRL}, the procedure generates a task offloading strategy for miners based on system states, e.g., blockchain transaction data, and observes the system reward at each time slot so that the offloading policy can be optimized (lines 6-12). Then the procedure updates the history experience tuple and train the Q-network (lines 14-18) with loss function minimization. This trial and error solution will avoid the requirement of prior information of offloading environment. Over the training time period, the trained neural network can characterize well the environment and therefore, the proposed offloading algorithm can dynamically adapt to the real mobile blockchain environment. 

\section{Simulation and Performance Evaluation }
\label{Section:Simulation}
 In this section, we investigate the proposed offloading scheme by conducting both real experiments and simulations evaluate the efficiency of computation offloading.
 \subsection{Implementation Settings}
 \subsubsection{{Experiment settings}}
\begin{figure}
 	\centering
 	\includegraphics[width=0.95\linewidth]{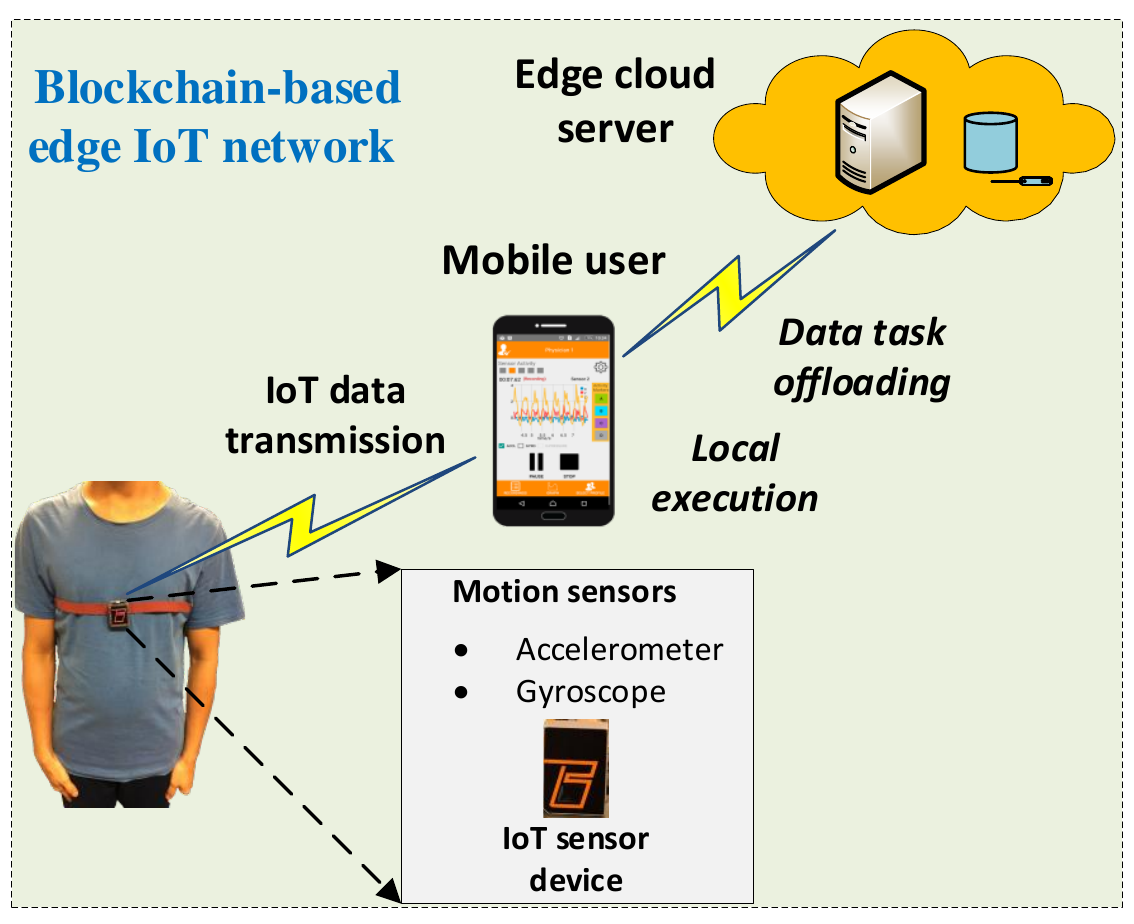}
 	\caption{ {Experiment setup for task offloading in blockchain.} }
 	\label{Experiment}
 	\vspace{-0.15in}
 \end{figure}

{ We considered a task offloading framework for the MEC-based blockchain network as shown in Fig.~\ref{Experiment}. We deployed an Ethereum blockchain network supported by Amazon cloud where edge computing was deployed on the Lambda Edge\footnote{https://aws.amazon.com/lambda/edge/} service for high transfer speeds and low-latency computation. A virtual machine Ubuntu 16.04 LTS, 2.4 GHz frequency was used as a computing unit for edge service. To test the proposed offloading scheme, we used a Sony mobile phone running on an Android OS version 8.0 platform with Qualcomm Snapdragon 845 processor, 4GB of RAM and 64GB of expandable storage, and a battery capacity of 2870mAh. The mobile devices connect with the edge cloud computing on the wireless network via Wi-Fi wireless communication with the standard IEEE 802.11g. A blockchain client was also installed on the device to transform the mobile phone into a miner node \cite{3}. }
 
 For IoT data generation, we used Biokin sensors \cite{3} as IoT devices to collect health data which then needs to be computed for medical services. For the evaluation of local task computation on smartphones, we employed the Firebase Performance Monitoring\footnote{https://firebase.google.com/docs/perf-mon} service to measure processing time and battery consumption. Meanwhile, for the evaluation of edge computing, we utilized the Kinesis Data Analytics service to monitor data streaming from mobile devices, and measure data computing latency and energy consumption. The evaluation needs IoT data and programming code for computation. We collected simultaneously data from sensors and stored them in separate files which are transmitted wirelessly to the mobile phone for computation.  Details of hardware settings for our offloading can be found in our recent works \cite{3, 41}. 
\subsubsection{Simulation settings}
In our simulation, a mobile blockchain network is considered with a MEC server with a varying number of mobile users (miners) and data processing tasks. The simulation is conducted over $T = 8,000$ timeslots, and each timeslot \textit{t} lasts 1 s. The computational capacity of the MEC server $F$ is set to 10 GHz/sec. At each mobile user, we set the local computation time and computing energy consumption as $4.75*10^{-7}$ s/bit and $3.25*10^{-7}$ J/bit, respectively \cite{9}. We assume that the size of data processing tasks  generated from IoT devices is randomly distributed between 50 kB and 150 kB \cite{36}. The local computation workload $X$ is set to 18000 CPU cycles/bit, and the delay threshold $\tau$ is assumed as 15s. The energy consumption efficiency coefficient and static circuit power of MEC server are set to $10^{-26}$ and 0.05W, respectively \cite{22}, and the channel gain factor $\sigma$ is set to 0.8. {Further, based on \cite{28, 29}, the channel state transition probability is set to $Pr(g^{t+1} =1|g^t=1) = Pr(g^{t+1} =0|g^t=0) = 0.95$.}

{For the mining tasks, we configure the simulation parameters according to \cite{36,40,39}. The size of block data to be mined is [5-10] kB. It is feasible in our blockchain scenario since data in the blockchain only includes metadata, while actual IoT data is stored in blockchain-basedd decentralized storage such as InterPlanetary File System (IPFS) \cite{3}. The hash power $p_n$ that the MU $n$ purchases from the edge-cloud provider follows a uniform distribution on [20-100] MHash/s. The hash power of the blockchain network follows a uniform distribution on $10^3$-$10^5$ GHash/s. Furthermore, the miner that first solves the cryptographically hard puzzles and achieves successfully consensus is rewarded with $\mathcal{R} = 30$ tokens \cite{36}.}

\begin{figure}
	\centering
	\includegraphics[width=0.95\linewidth]{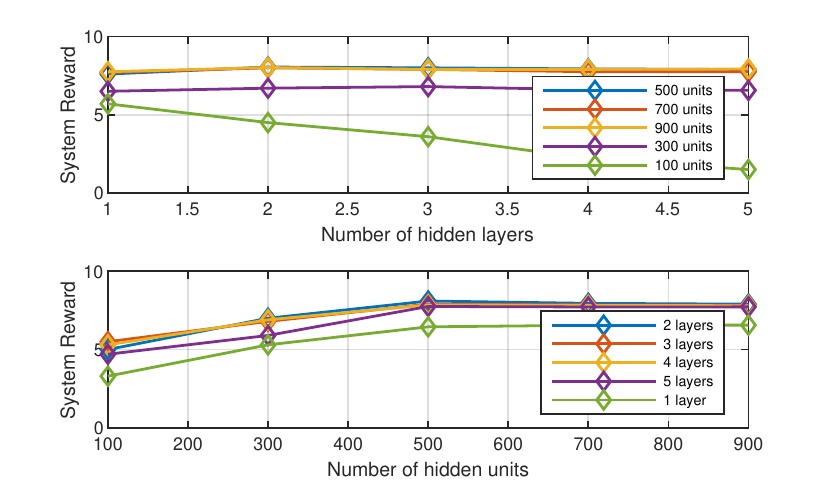}
	\caption{{Impacts of different hyperparameters.} }
	\label{learningSetting}
	\vspace{-0.2in}
\end{figure}
\begin{figure*}[t!]
	\centering
	\begin{subfigure}[t]{0.5\textwidth}
		\centering
		\includegraphics[width=0.95\linewidth]{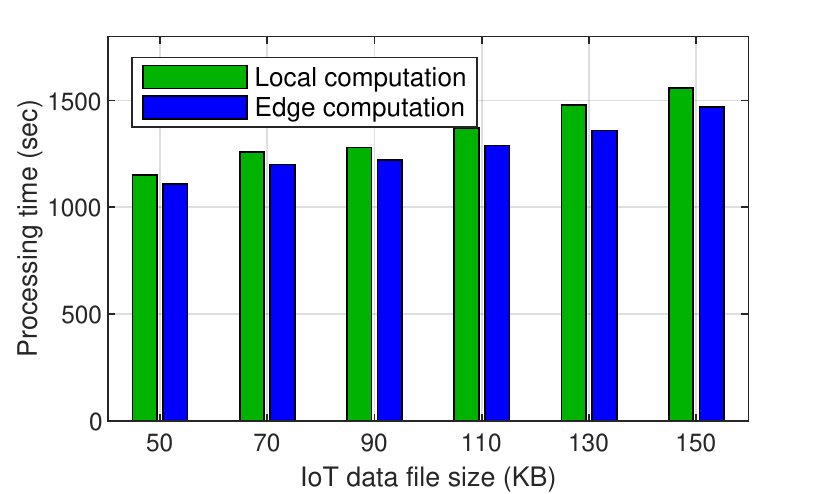} 
		\caption{Processing time. }
	\end{subfigure}%
	~
	\begin{subfigure}[t]{0.5\textwidth}
		\centering
		\includegraphics[width=0.95\linewidth]{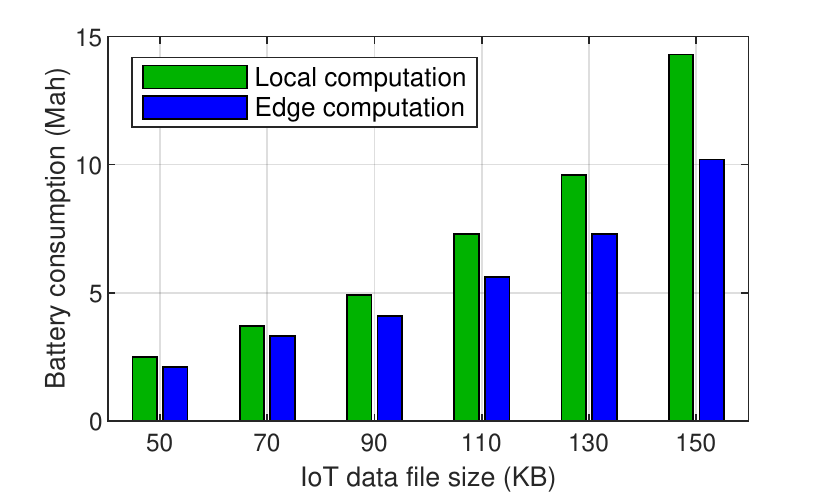} 
		\caption{Battery consumption.}
	\end{subfigure}
	\caption{Experimental results for local and edge computation.}
	\label{experimental_Result}
	\vspace{-0.2in}
\end{figure*}

\begin{figure*}[t!]
	\centering
	\begin{subfigure}[t]{0.5\textwidth}
		\centering
		\includegraphics[width=0.95\linewidth]{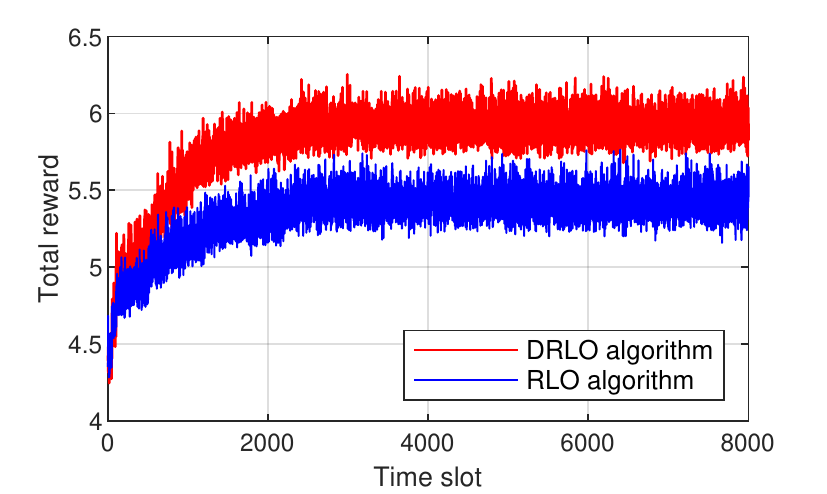} 
		\caption{Convergence performance with $\beta = 0.5$. }
	\end{subfigure}%
	~ 
	\begin{subfigure}[t]{0.5\textwidth}
		\centering
		\includegraphics[width=0.95\linewidth]{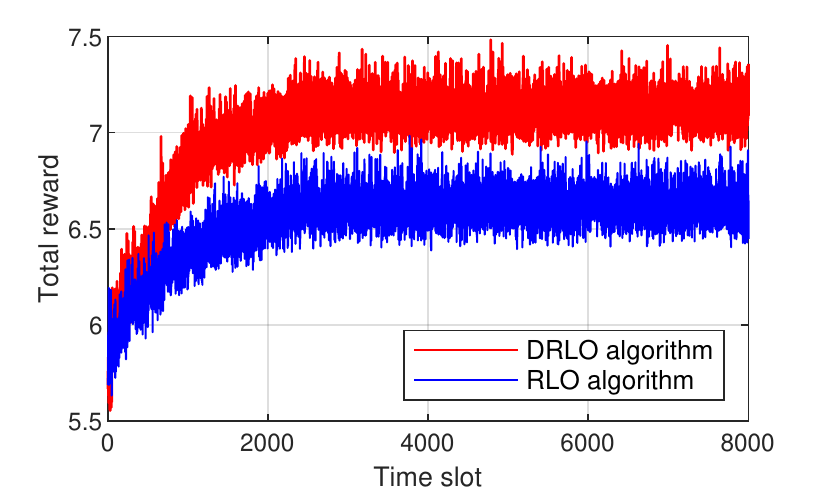} 
		\caption{Convergence performance with $\beta = 0.8$.}
	\end{subfigure}
	\caption{Convergence performance of offloading algorithms with $\beta = 0.5$ and with $\beta = 0.8$.}
	\label{Convergence_performance}
	\vspace{-0.2in}
\end{figure*}

In the DQN-learning algorithm, we configure the parameters as follows. The discount factor $\gamma$ is set to 0.85; the learning rate equals 0.01, and the replay memory capacity and training batch size are set to $10^5$ and $128$, respectively. {We have employed gradient descent when approximating RL, and used a feedforward neural network to build our DNN model.} Next, we employ ReLU as the activation function in the hidden layers, while the sigmoid activation function is utilized in the output layer to relax the offloading decision variables \cite{41}.

{The architecture of the deep neural network significantly affects the performance of the deep reinforcement learning algorithms, and thus it requires a thoughtful design. We implemented simulations to select the parameter for our learning structure with the different DNN layers and varying hidden units. The experiment results in Fig.~\ref{learningSetting} assert that the number of hidden layers and neurons have a direct effect on the system throughput (e.g., system reward). Increasing the layers leads to better results, but at some point when the network is made deeper, the results start degrading. Their result shows that when using two hidden layers with 500 hidden units (including 300 units in first hidden layer and 200 units in second hidden layers), the model achieves the best performance. Hence, we choose a two-layer 500 DNN to build the offloading learning optimizer in this work.}

To evaluate the task offloading performance for mobile blockchain, we focus on the following three metrics.
\begin{itemize}
	\item The computation latency for offloading and local execution. 
	\item The energy consumed by mining process on MEC server and local computation.
	\item The privacy level during the data processing task offloading process.
\end{itemize}
To highlight the advantage of the proposed offloading schemes in terms of latency and energy metrics, we compare our RLO and DRLO algorithms with two baseline schemes, i.e., 
\begin{itemize}
	\item Non-offloading scheme (NO): All data processing tasks are executed at the local devices, (.i.e, setting offloading decision vector $ x_{nm}=0 (\forall \textit{n} \in \mathcal{N},  \textit{m}  \in \mathcal{M} $)).
	\item Edge offloading scheme (EO):  All miners offload their data processing tasks to the MEC server (.i.e, setting offloading decision vector $ x_{nm}=1 (\forall \textit{n} \in \mathcal{N},m \in \mathcal{M} $)).
\end{itemize}

Futher, to evaluate the privacy performance metric, we compare our algorithms with the constrained Markov decision process (CMDP)-based scheme \cite{28} and RL-based design in \cite{29}.

Specially, we introduce a tradeoff factor $\beta \in [0, 1]$ for each miner to analyze the influence of computation laytency and energy consumption on the quality of mobile blockchain offloading. In this way, the weighted factors in equation~\ref{Equation:SystemCost} can be represented by $\alpha_1 = \beta$ and $\alpha_2 = 1- \beta$. Therefore, the reward function equation~\ref{Equation:Reward} can be rewritten as 
\begin{equation}
r^t(s,a) = P^t(s,a) - [\beta E^t(s,a) + (1-\beta) L^t(s,a)], 
\end{equation}
where $E^t(s,a)$ and $L^t(s,a)$ are the sum offloading energy and latency cost, respectively.

\subsection{{Experimental Evaluation}}
\begin{figure*}[t!]
	\centering
	\begin{subfigure}[t]{0.3\textwidth}
		\centering
		\includegraphics[width=0.95\linewidth]{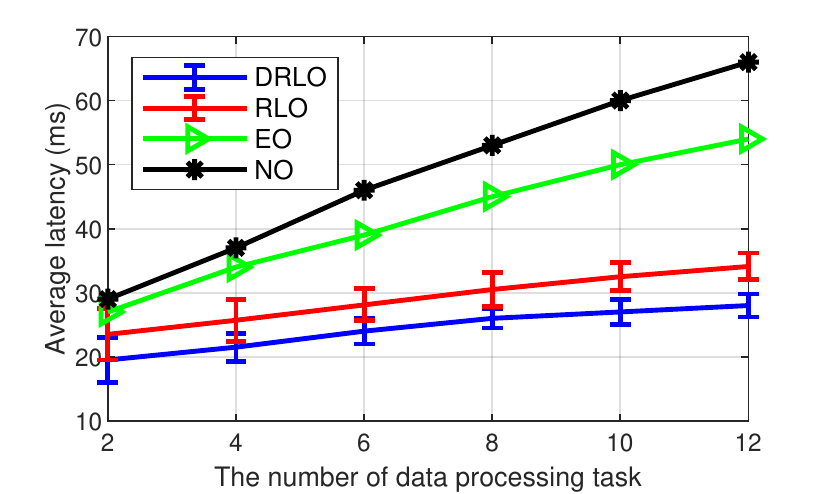} 
		\caption{Average latency. }
	\end{subfigure}%
	~
	\begin{subfigure}[t]{0.3\textwidth}
		\centering
		\includegraphics[width=0.95\linewidth]{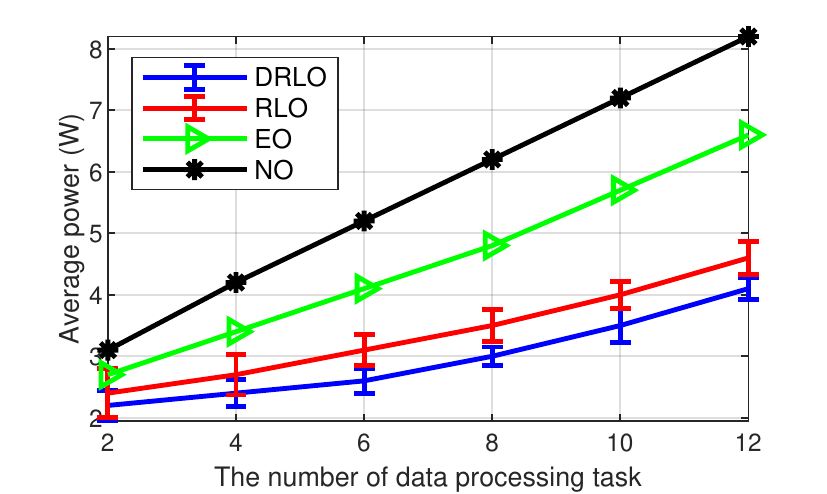} 
		\caption{Average power.}
	\end{subfigure}
	~
	\begin{subfigure}[t]{0.3\textwidth}
		\centering
		\includegraphics[width=0.95\linewidth]{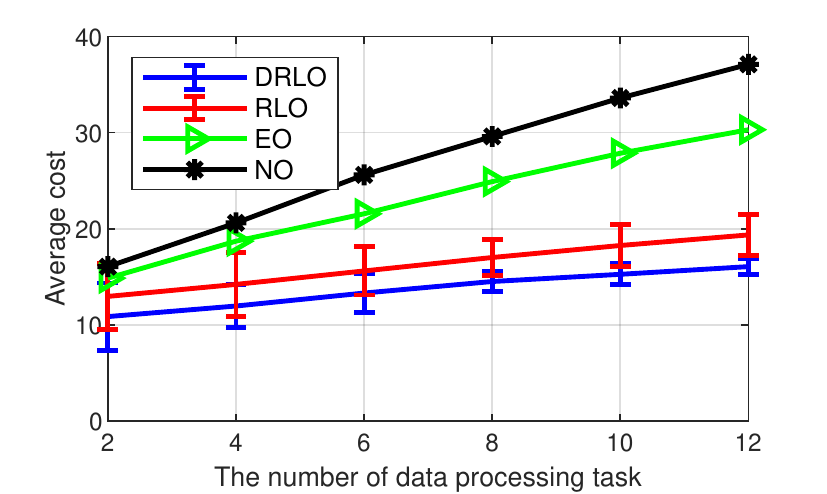} 
		\caption{Average offloading cost.}
	\end{subfigure}
	\caption{Comparison results for single user scenario with $\beta = 0.5$.}
	\label{OffloadingPerformance_05}
	\vspace{-0.2in}
\end{figure*}

\begin{figure*}[t!]
	\centering
	\begin{subfigure}[t]{0.3\textwidth}
		\centering
		\includegraphics[width=0.95\linewidth]{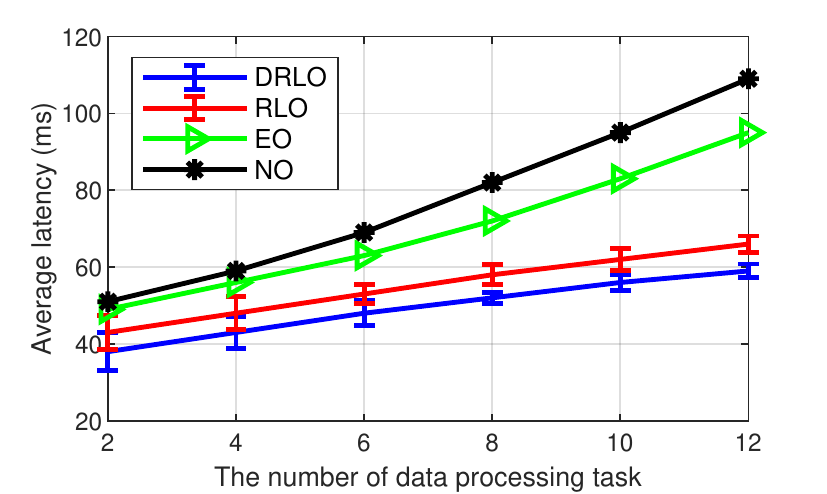} 
		\caption{Average latency. }
	\end{subfigure}%
	~
	\begin{subfigure}[t]{0.3\textwidth}
		\centering
		\includegraphics[width=0.95\linewidth]{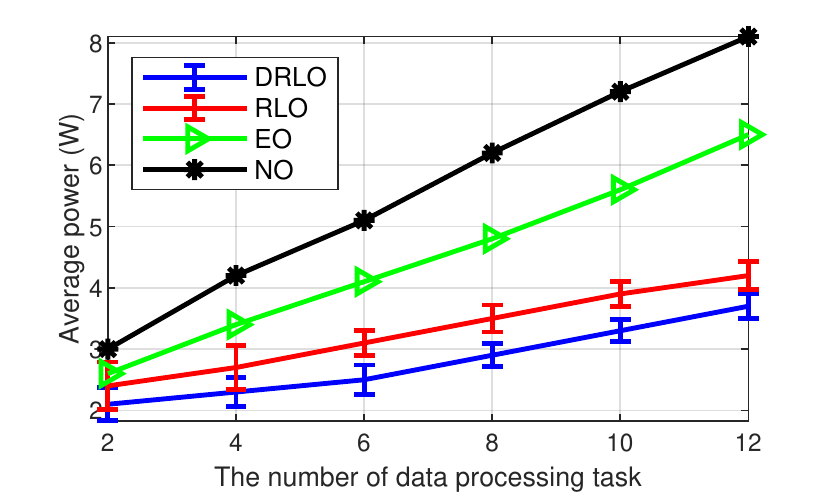} 
		\caption{Average power.}
	\end{subfigure}
	~
	\begin{subfigure}[t]{0.3\textwidth}
		\centering
		\includegraphics[width=0.95\linewidth]{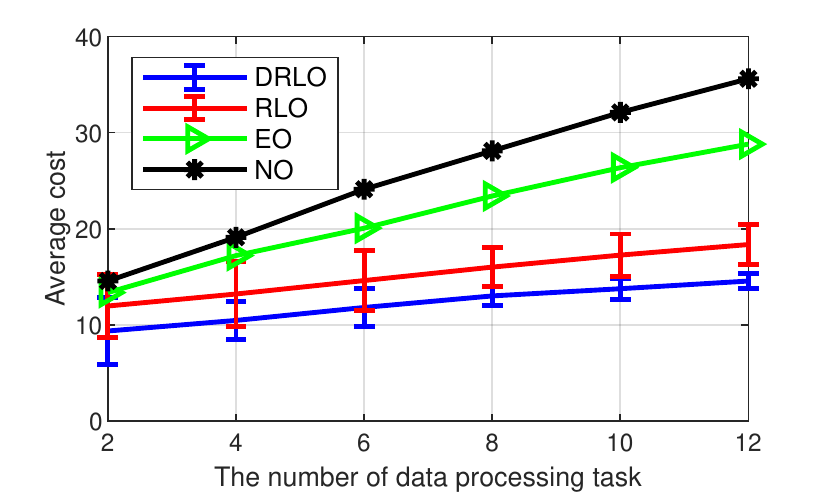} 
		\caption{Average offloading cost.}
	\end{subfigure}
	\caption{Comparison results for single user scenario with $\beta = 0.8$.}
	\label{OffloadingPerformance_08}
	\vspace{-0.2in}
\end{figure*}
{We considered two computation scenarios to prove the system efficiency: local execution and offloading to the edge computing service for computing IoT data. In the first scenario, the IoT data are executed locally on the mobile device. In the second scenario, the IoT data are offloaded for execution on the edge computing using our settings. A set of sensor data files with different sizes (50 kB-150 kB) was used to evaluate the proposed scheme. Each IoT data input is executed 10 times, and the average values are obtained. 
}

We investigated the implementation results via two performance metrics: processing time and energy consumption as shown in Fig.~\ref{experimental_Result}. The processing time includes execution time, and offloading time with downloading time in the case of task offloading to the edge cloud server. In Fig.~\ref{experimental_Result}(a), the average processing time of local computation is higher than that of edge computation for each IoT data file size. For example, the offloading scheme can save 36\% time for completing computation of a 50 kB file and save up to 43\% time for computing a 150 kB file, which shows advantages of the offloading scheme. 

In the experimental result for battery consumption in Fig.~\ref{experimental_Result}(b), IoT data tasks consumes less energy when being executed with the offloading scheme because the resource-intensive computational tasks are offloaded to the edge-cloud servers. As an example, offloading the 50 kB file consumes less 15\% energy than the case of local computation. Specially, the energy usage of the offloading scheme become more efficient when the data size increases. For instance, executing a 130 kB file and 150 kB file can save 17.8\% and 19\% energy, respectively when offloading the task to the edge-cloud server, compared to local execution.

\subsection{{Simulation Evaluation}}
\subsubsection{Convergence performance}
We first evaluate the convergence performance of our data processing task offloading algorithms, namely RLO and DRLO with tradeoff factor settings $\beta =0.5$ and $\beta =0.8$ through the training process. Fig.~\ref{Convergence_performance} shows the learning curve obtained by the offloading policy using the proposed algorithms over 8,000 timeslots. The results show that the total system reward is very small at the beginning of the learning process. However, as the number of timeslot increases, the total reward increases rapidly and become stable after about 2500 timeslots, which validates the convergence performance of the proposed DQN-learning scheme. Further, it can be seen that the DRLO algorithm can achieve a better long-term reward in both cases, compared to the RLO scheme. For instance, in case of $\beta =0.5$ in Fig.~\ref{Convergence_performance}(a), the DRLO-based learning strategy converges to 5.9, which is approximately 5.2\% higher than that of the RLO-based learning scheme, which converges to 5.4 after about 2500 timeslots. Moreover, for a larger tradeoff value with $\beta =0.8$ in Fig.~\ref{Convergence_performance}(b), the DRLO scheme converges to about 7.1, which is roughly 9\% higher than that of the RLO scheme. Overall, the DRLO scheme achieves better performance for all cases. The key reason behind that is the RLO with Q-learning always chooses system actions in a greedy manner. On the contrary, DRLO with deep Q-learning applies trial-and-error search to balance exploration and exploitation. {In addition, the DRLO scheme with a DNN as an approximator outperforms in obtaining more valuable features of certain parts of states, by using the iterative target network. This also prevents the instabilities to propagate quickly and reduces the risk of divergence. As a result, the DRLO scheme can learn better from the state space for making better decisions which leads to an improvement of system reward.} Next, we verify the proposed schemes via single user and multiple user scenarios. The simulation results are averaged from 50 runs of numerical simulations. 

\subsubsection{Task offloading in single user scenario}

In the single user model, we consider a blockchain network with a single miner \textit{N} = 1 and change the number of tasks at each miner (\textit{M}=2$\sim$12). Note that the size of each task is the same for all miners. The simulation results of the offloading performance in case of $\beta =0.5$ and $\beta =0.8$ are shown in Fig.~\ref{OffloadingPerformance_05} and Fig.~\ref{OffloadingPerformance_08}. In Fig.~\ref{OffloadingPerformance_05} with $\beta =0.5$, it is observed that when the number of task increases, the cost of four approaches increases due to the growing size of transaction data on the blockchain network. For example, in the EO strategy, the average offloading cost of all miners increases from 15 at $M= 2$ to 30 at $M=12$, and that of the RLO scheme also increases to 9 at $M=12$. Further, the NO algorithm always incurs a higher energy cost than the other schemes and has the largest increasing rate of 125\%, reaching 36 at $M=12$. The reason behind this observation is that when the number of data processing task grows, the computational capacity of the mobile miner becomes less sufficient to provide mining services for computing all tasks. Thus newly generated blockchain transactions have to wait to be processed in the buffer of local devices, which thus increases the mining latency. Further, the higher cost comes from a higher computation delay and a power consumption. Specially, the DRLO scheme achieves the best performance with minimum sum cost for all data processing tasks. 

Fig.~\ref{OffloadingPerformance_08} shows the simulation results with a larger tradeoff factor $\beta =0.8$. According to the formulated reward function (18), a larger tradeoff factor will give more penalty to energy consumption. In this case, the gap between the DRLO scheme and the other baselines is larger than that in the case of $\beta =0.5$, which is caused by the increased offloading delay and lower power cost. Particularly, the DRLO algorithm exhibits the best performance again among four schemes with minimum offloading efficiency index. For instance, when the miner has 12 data processing tasks, the offloading cost averaged over 50 simulations of the DRLO scheme is 18.7\%, 57\%, and 65\% lower than those of the RLO, EO, and NO schemes, respectively. Such results demonstrate the efficiency of the DRLO-based scheme in the data task offloading.

\subsubsection{Task offloading in multi-user scenario}
Next, we analyze the task offloading performance for the blockchain network with multiple users. We consider two cases, $N=5$ and $N=10$. We also assume that there are 2 data processing tasks at each miner, and the size of each task is the same for all miners. The comparison results are shown in Table~\ref{Table:Offloading_MultiUser_05} and Table~\ref{Table:Offloading_MultiUser_08} for tradeoff factor $\beta = 0.5$ and $\beta = 0.8$, respectively. Under the scenario with $\beta = 0.5$ in Table~\ref{Table:Offloading_MultiUser_05}, the DRLO scheme exhibits the lowest average power consumption, computation latency in both cases $N=5$ and $N=10$, and thus achieves the minimum total offloading cost, followed by the RLO scheme with a small gap. Another observation is that among the four schemes, the highest offloading cost comes from the EO-based scheme, instead of the NO-based scheme. This is because the more miners offload data processing tasks to MEC server, the more computing capacity is required to provide enough computation resources for running the data processing tasks of all blockchain users. It is also noteworthy that the computation capacity of the MEC server is only sufficient to provide resources for a certain number of miners and high mining demands from multiple users can result in a significant increase in network latency and system cost. 

\begin{table}
	\scriptsize
	\centering
	\captionsetup{font=scriptsize}
	\caption{\textbf{Comparison results for multi-user scenario with $\beta =0.5$.}}
	\label{Table:Offloading_MultiUser_05}
	
	\begin{tabular}{|c||c c|c c|c c|}
		\hline
		\multirow{2}{*}{Schemes} &
		\multicolumn{2}{c|}{Average power (W) } &
		\multicolumn{2}{c|}{Average latency(ms)} &
		\multicolumn{2}{c|}{Average cost} \\
		& \textit{N=5} & \textit{N=10} &\textit{N=5} & \textit{N=10}  & \textit{N=5} & \textit{N=10} \\
		\hline
		DRLO & \textbf{3.5} & \textbf{6.9} & \textbf{32.6 }& \textbf{55.8} & \textbf{19.8}& \textbf{34.8} \\
		
		RLO & 4.2 & 8.6 & 35.3 & 63.2 & 21.9 & 40.2 \\
		
		EO & 6.2 & 14.3 & 53.2 & 94.6 & 32.8 & 53.5 \\
		
		NO & 4.5 & 9.1 & 39.7 & 75.4 & 24.6 & 46.8 \\
		\hline
	\end{tabular}
\end{table}
\begin{table}
	\centering
	\scriptsize
	\captionsetup{font=scriptsize}
	\caption{\textbf{Comparison results for multi-user scenario with $\beta =0.8$.}}
	\label{Table:Offloading_MultiUser_08}
	
	\begin{tabular}{|c||c c|c c|c c|}
		\hline
		\multirow{2}{*}{Schemes} &
		\multicolumn{2}{c|}{Average power (W) } &
		\multicolumn{2}{c|}{Average latency(ms)} &
		\multicolumn{2}{c|}{Average cost} \\
		& \textit{N=5} & \textit{N=10} &\textit{N=5} & \textit{N=10}  & \textit{N=5} & \textit{N=10} \\
		\hline
		DRLO & \textbf{3.3} & \textbf{6.5} & \textbf{55.9 }& \textbf{90.7} & \textbf{13.8}& \textbf{23.3} \\
		
		RLO & 3.9 & 8.3 & 59.1 & 98.5 & 14.9 & 26.3 \\
		
		EO & 6.0 & 13.9 & 82.2 & 163.8 & 21.2 & 35.6 \\
		
		NO & 4.2 & 8.5 & 66.3 & 116.7 & 16.6 & 30.1 \\
		\hline
	\end{tabular}
\vspace{-0.15in}
\end{table}

Using a higher tradeoff value $\beta = 0.8$, we present comparison results in Table~\ref{Table:Offloading_MultiUser_08}. Similar to the single user scenario, the larger tradeoff factor gives more penalty to energy consumption, and therefore, the energy consumed by all miners is lower than that of the case of $\beta = 0.5$, while the offloading delay becomes larger for both cases of $N=5$ and $N=10$. Based on such observations, we can minimize energy consumption with respect to task offloading latency by adjusting the tradeoff factor for a better offloading efficiency. It is also worth mentioning that the DRLO-based scheme still exhibits the best offloading performance with minimum power usage, mining delay and resulting offloading cost. For instance, in the case of $N=10$, the offloading cost of the DRLO-based strategy is 12.7\%, 52.5\%, and 30.4\% lower than those of the RLO, EO, and NO schemes, respectively. The numerical results clearly show that the DRLO algorithm outperforms the other baseline schemes in improving offloading cost efficiency in multi-user scenarios. 
\begin{figure}
	\centering
	\includegraphics[width=0.95\linewidth]{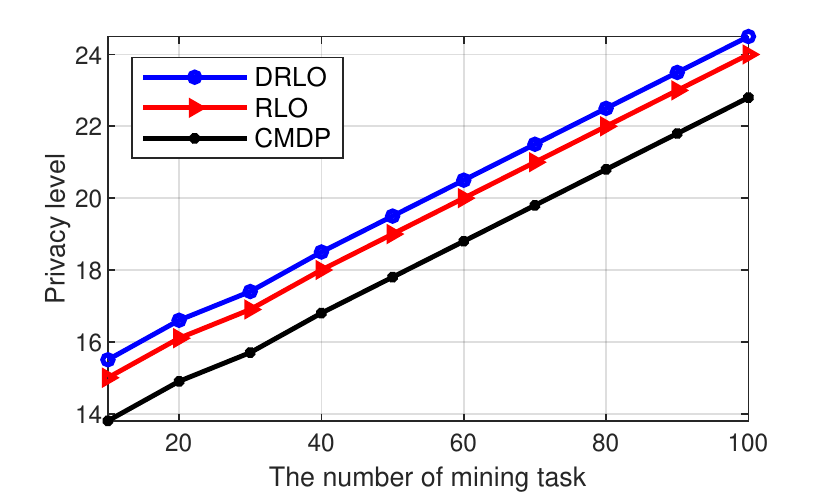}
	\caption{The achieved privacy level.}
	\label{Privacy_Performance}
	\vspace{-0.2in}
\end{figure}
\subsubsection{Privacy performance}

Finally, we analyse the performance of the proposed offloading design in terms of the privacy metric as shown in Fig.~\ref{Privacy_Performance}. We compare the privacy performance of our DRLO-based scheme with the RL-based scheme \cite{28} and CMDP-based scheme \cite{29}. It can be observed from Fig.~\ref{Privacy_Performance} that the privacy level of the blockchain miner increases when the amount of transaction data increases from 10 kB to 100 kB in each time slot for all offloading schemes. However, the proposed DRLO method can achieve the best privacy performance, compared to the other benchmarks. For instance, as the mobile user mines 10 kB blockchain transactions, the proposed DRLO scheme achieves 5.2\% and 12.7\% higher privacy levels compared with the RL-based and CMDP-based schemes, respectively. Further, in the case of mining 100 kB blockchain transactions, the privacy level of DRLO scheme still shows the best performance, with 5.5\% higher than the RL-based strategy and 13.4\% higher than the CMDP-based strategy. 

\subsection{{Discussion and Future Works}}
{In this paper, we proposed a DRL-based dynamic task offloading scheme for a MEC blockchain network. Our focus was on formulating task offloading and user privacy preservation as a joint optimization problem. We then developed a RL-based algorithm to solve the proposed optimization problem with a simplified blockchain model. To this end, we conducted both experiments and numerical simulations to verify the effectiveness of the proposed offloading algorithm in terms of various performance metrics and compare with other offloading schemes. The simulation results with our proposed schemes showed the performance of computation offloading for mobile blockchain networks can be significantly improved in terms of enhanced privacy level and reduced system costs, compared with other blockchain offloading schemes.}

{As future works, we will extend the proposed DRL scheme by taking action space variation into consideration. In this paper, we assume that the action space is stable during the learning process, but in certain blockchain scenarios, the action set can vary over time due to the dynamicity of the MEC-based blockchain offloading system, i.e., dynamic user demands, dynamic MEC resource usage, dynamic wireless channel power usage. In this regard, developing adative and online deep RL schemes would be helpful to cope with the dynamic environment. Another direction is to consider the dynamics of the network with the variable user set. In this case, the number of users in our blockchain network might vary with time, i.e., the current users can exit the network and new users can join the network. To cope with the dynamics of the network, one solution can be applied as follows. We can set a training value so that the remaining user states in the DNN network are regarded as empty states if the number of users is smaller than that preset training value. For instance, assuming that the preset training value is $X_p$ and the number of current users is $X_u (X_p> X_u)$, we can set the remaining $X_p- X_u$ states to be 0 and drop the remaining $X_p- X_u$  actions in the output of the DRLO algorithm. On the other hand, if  $X_p< X_u$, we portion the collected  user states into several parts and input these parts to the actor network in turn. This solution enables flexible adjustment on the learning algorithm according to the dynamics of the network \cite{36}. }
\section{Conclusions}
\label{Section:Conclusions}
In this paper, we have proposed reinforcement learning-based task offloading algorithms for multi-users to obtain the optimal offloading policy in a dynamic blockchain network with MEC. We have formulated the task offloading and privacy preservation as a joint optimization problem. A RL-based scheme using the Q-network algorithm is employed to learn efficiently the offloading policy such that the total system cost combining computation latency and energy consumption is minimized while guaranteeing the best user privacy and {mining reward} performance. To break the curse of high dimensionality in state space, we have then developed a DRL-based approach using a deep Q-network algorithm. The offloading performances in terms of energy consumption, computation latency, and user privacy are analyzed under various conditions for both single user and multiple user offloading scenarios via numerical simulations. The experimental results have clearly showed that the proposed DRL offloading scheme is superior to the other baseline methods with a reduced energy consumption, computation latency with much lower offloading costs and improved privacy level.

\bibliography{Ref}
\bibliographystyle{IEEEtran}
\begin{IEEEbiography}[{\includegraphics[width=1in,height=1.25in,clip,keepaspectratio]{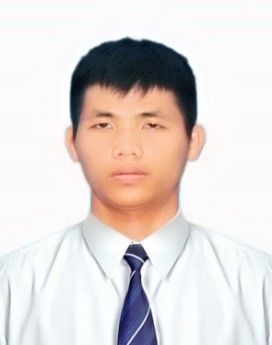}}]{Dinh C. Nguyen}
(Graduate Student Member, IEEE) is currently pursuing the Ph.D. degree at the School of Engineering, Deakin University, Victoria, Australia. His research interests focus on blockchain, deep reinforcement learning, mobile edge/cloud computing, network security and privacy. He is currently working on blockchain and reinforcement learning for Internet of Things and 5G networks. He has been a recipient of the prestigious Data61 PhD scholarship, CSIRO, Australia.
\end{IEEEbiography}

\begin{IEEEbiography}[{\includegraphics[width=1in,height=1.25in,clip,keepaspectratio]{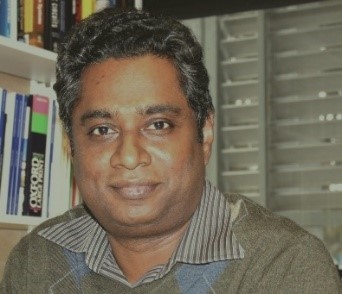}}]{Pubudu N. Pathirana}
(Senior Member, IEEE) was born in 1970 in Matara, Sri Lanka, and was educated at Royal College Colombo. He received the B.E. degree (first class honors) in electrical engineering and the B.Sc. degree in mathematics in 1996, and the Ph.D. degree in electrical engineering in 2000 from the University of Western Australia, all sponsored by the government of Australia on EMSS and IPRS scholarships, respectively. He was a Postdoctoral Research Fellow at Oxford University, Oxford, a Research Fellow at the School of Electrical Engineering and Telecommunications, University of New South Wales, Sydney, Australia, and a Consultant to the Defence Science and Technology Organization (DSTO), Australia, in 2002. He was a visiting professor at Yale University in 2009. Currently, he is a full Professor and the Director of Networked Sensing and Control group at the School of Engineering, Deakin University, Geelong, Australia and his current research interests include Bio-Medical assistive device design, human motion capture, mobile/wireless networks, rehabilitation robotics and radar array signal processing.
\end{IEEEbiography}

\begin{IEEEbiography}[{\includegraphics[width=1in,height=1.25in,clip,keepaspectratio]{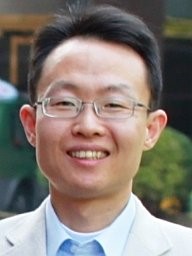}}]{Ming Ding}
(Senior Member, IEEE) received the B.S. and M.S. degrees (Hons.) in electronics engineering and the Ph.D. degree in signal and information processing from Shanghai Jiao Tong University (SJTU), Shanghai, China, in 2004, 2007, and 2011, respectively. From April 2007 to September 2014, he worked as a Researcher/Senior Researcher/Principal Researcher at the Sharp Laboratories of China, Shanghai. He also served as the Algorithm Design Director and the Programming Director for a system-level simulator of future telecommunication networks in Sharp Laboratories of China for more than seven years. He is currently a Senior Research Scientist with the CSIRO Data61, Sydney, NSW, Australia. His research interests include information technology, data privacy and security, machine Learning and AI. He has authored over 100 articles in IEEE journals and conferences, all in recognized venues, and around 20 3GPP standardization contributions, and a Springer book Multi-Point Cooperative Communication Systems: Theory and Applications. He holds 21 U.S. patents and co-invented another more than 100 patents on 4G/5G technologies in CN, JP, KR, EU. He is an Editor of the IEEE TRANSACTIONS ON WIRELESS COMMUNICATIONS and the IEEE Wireless Communications Letters. Besides, he is or has been a Guest Editor/CoChair/Co-Tutor/TPC Member of several IEEE top-tier journals/conferences, such as the IEEE JOURNAL ON SELECTED AREAS IN COMMUNICATIONS, IEEE Communications Magazine, and the IEEE GLOBECOM Workshops. He was the Lead Speaker of the industrial presentation on unmanned aerial vehicles in IEEE GLOBECOM 2017, which was awarded as the Most Attended Industry Program in the conference. He was awarded as the Exemplary Reviewer of the IEEE TRANSACTIONS ON WIRELESS COMMUNICATIONS in 2017.
\end{IEEEbiography}

\begin{IEEEbiography}[{\includegraphics[width=1in,height=1.25in,clip,keepaspectratio]{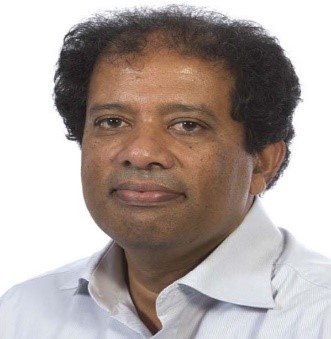}}]{Aruna Seneviratne}
 (Senior Member, IEEE) is currently a Foundation Professor of telecommunications with the University of New South Wales, Australia, where he holds the Mahanakorn Chair of telecommunications. He has also worked at a number of other Universities in Australia, U.K., and France, and industrial organizations, including Muirhead, Standard Telecommunication Labs, Avaya Labs, and Telecom Australia (Telstra). In addition, he has held visiting appointments at INRIA, France. His current research interests are in physical analytics: technologies that enable applications to interact intelligently and securely with their environment in real time. Most recently, his team has been working on using these technologies in behavioral biometrics, optimizing the performance of wearables, and the IoT system verification. He has been awarded a number of fellowships, including one at British Telecom and one at Telecom Australia Research Labs.
\end{IEEEbiography}

\end{document}